\documentclass[
    reprint,
    superscriptaddress,
    nofootinbib,
    amsmath,amssymb,
    aps,
    floatfix
]{revtex4-1}

\usepackage[utf8]{inputenc}

\usepackage{graphicx}
\usepackage{dcolumn}
\usepackage{multirow}
\usepackage{placeins}

\usepackage{bm}
\usepackage{upgreek}
\usepackage{commath}
\usepackage[version=4]{mhchem}

\usepackage{xcolor}

\usepackage[hidelinks]{hyperref}

\begin{document}

\title{Universal energy barrier for plastic flow in a network glass}

\author{Ibrahim~Ghanem}
\affiliation{Department of Microsystems Engineering, University of Freiburg, Georges-K\"ohler-Allee 103, 79110 Freiburg, Germany}

\author{Richard~Jana}
\affiliation{Department of Microsystems Engineering, University of Freiburg, Georges-K\"ohler-Allee 103, 79110 Freiburg, Germany}
\affiliation{Department of Chemistry and Materials Science, Aalto University, 02150 Espoo, Finland}

\author{Wolfram~G.~Nöhring}
\affiliation{Department of Microsystems Engineering, University of Freiburg, Georges-K\"ohler-Allee 103, 79110 Freiburg, Germany}

\author{Michael Moseler}
\affiliation{Fraunhofer IWM, MicroTribology Center $\upmu$TC, W\"ohlerstra\ss e 11, 79108 Freiburg, Germany}
\affiliation{Institute of Physics, University of Freiburg, Hermann-Herder-Stra\ss e 3, 79104 Freiburg, Germany}

\author{Lars Pastewka}
\email{lars.pastewka@imtek.uni-freiburg.de}
\affiliation{Department of Microsystems Engineering, University of Freiburg, Georges-K\"ohler-Allee 103, 79110 Freiburg, Germany}

\date{\today}

\begin{abstract}
Shear transformations are the elementary events of plasticity in glasses and underpin many mesoscopic models of plastic flow, yet their atomic origin and energetics remain difficult to establish.
Here we use a machine-learned interatomic potential to show that, in amorphous carbon as a representative network glass, a shear transformation corresponds to the breaking or formation of a single covalent bond.
The simplicity of the associated reaction coordinate, the bond length, enables us to extract energy landscapes and barriers.
Despite the intrinsic structural disorder, we find that all energy landscapes collapse onto a universal barrier shape.
What remains bond-specific is a single scale, the bond-jump distance, which fixes the barrier height through a cubic law.
Combined with the activation statistics of shear transformations, this yields a model that predicts the plastic flow stress from independently determined bond-level quantities.
Our results provide microscopic support for phenomenological models of plastic flow in glasses and suggest a route towards a quantitative description of amorphous plasticity.
\end{abstract}

\maketitle

Crystalline solids deform plastically through the motion of dislocations~\cite{anderson_theory_2017}, but glasses lack the long-range order required to define such line defects.
Plastic deformation in amorphous solids instead proceeds through localized atomic rearrangements known as shear transformations (STs)~\cite{spaepen_microscopic_1977,argon_plastic_1979,falk_dynamics_1998}.
These elementary events underlie a broad class of mesoscopic descriptions, from shear-transformation-zone theory~\cite{falk_deformation_2011} to stochastic and elastoplastic lattice models~\cite{bulatov_stochastic_1994,talamali_strain_2012,Barrat_2018}.
Although such models successfully describe many collective features of amorphous plasticity, the local rules governing individual transformations are generally prescribed at the mesoscopic level rather than derived from well-defined plastic deformation units~\cite{Castellanos_2021}.
In widely studied model glasses, STs involve the cooperative motion of many atoms~\cite{tanguy_plastic_2006}, making it difficult to identify a simple microscopic reaction coordinate and isolate the associated energetics~\cite{rodney_distribution_2009,boioli_shear_2017,xu_strain_2017}.
This has limited attempts to connect the energetics of elementary STs quantitatively to the macroscopic flow response.

Covalent network glasses provide a natural setting in which to seek such a connection.
Their local structure is encoded by a discrete network of bonds, providing a direct description of changes in atomic connectivity.
In silica glasses, atomistic studies have linked elementary plastic rearrangements to bond breaking and changes in local network topology~\cite{Bonfanti2019,bamer-2025}.
Here, we investigate amorphous carbon (a-C)~\cite{Robertson_2002,Ohtake_2021} using the machine-learned atomic cluster expansion potential (ACE)~\cite{Qamar2023-nw}.

\begin{figure}
\centering
    \includegraphics[]{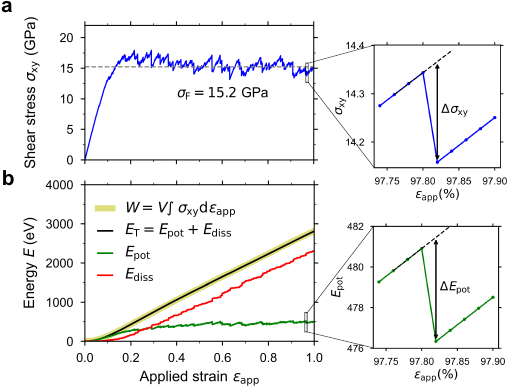}
    \caption{\textbf{Phenomenology of plastic flow.}
    (a) Stress–strain response of an a-C network at $2.7~\text{g}\,\text{cm}^{-3}$.
    The dashed gray line marks the mean flow stress ($15.2~\text{GPa}$).
    (b) Potential, dissipated, and total input energies versus strain.
    Insets define the plastic stress drop $\Delta\sigma_{\text{xy}}$ and potential energy drop $\Delta E_{\text{pot}}$, respectively.}
    \label{fig:phenomenology}
\end{figure}

We consider amorphous structures spanning mass densities $\rho$ from $2.7$ to $3.3~\text{g}\,\text{cm}^{-3}$, generated by a liquid-quench protocol and deformed by athermal quasistatic shear~\cite{Maeda1981-ay,Maloney2006-iy} (Supplementary Note~S-I).
Under shear, a-C shows a stress–strain response (Fig.~\ref{fig:phenomenology}a) similar to other model glasses~\cite{Malandro1999-th,Barrat_kob_stress_strain}.
The stress increases linearly with strain, peaks at yielding, then levels off at a nearly constant value~\cite{Jana2020-sc}.
In this plastic flow regime, the stress fluctuates and undergoes sudden drops that coincide with drops in the potential energy (Fig.~\ref{fig:phenomenology}a,b).
These serrations, caused by localized mechanical instabilities, are the characteristic signatures of STs.

To determine how mechanical work is partitioned during deformation, we track separately the stored potential energy $E_{\rm pot}$ and the dissipated energy $E_{\rm diss}$.
Their sum is the total energy input into the system, $E_T=E_{\rm pot}+E_{\rm diss}$, which coincides with the work $W$ obtained from the area under the stress--strain curve (Fig.~\ref{fig:phenomenology}b).
The dissipative contribution $E_{\rm diss}$ is negligible while the solid deforms elastically but becomes dominant once it yields.
In steady-state flow, $E_{\rm diss}$ grows approximately linearly with strain ($\sim 0.7$~eV per unit strain per atom), whereas $E_{\rm pot}$ changes only weakly ($\sim 0.03$~eV per unit strain per atom).
Consequently, nearly all mechanical work supplied during plastic flow is dissipated rather than stored.
This observation defines the program of this Article, namely that if the energetics and activation statistics of individual STs can be described microscopically, then both the macroscopic dissipation rate and the plastic flow stress follow from microscopic quantities alone.

\begin{figure*}
    \centering
    \includegraphics[]{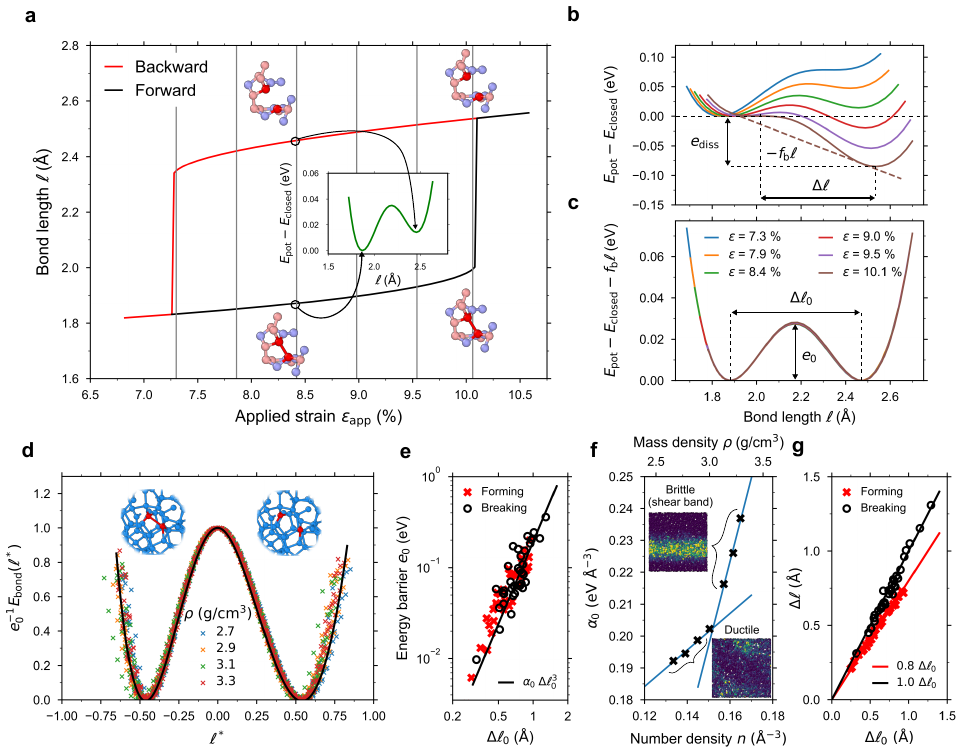}
\caption{{\bf The making and breaking of a single bond.}
            (a) Hysteresis loop of a reversible bond. Inset shows the potential energy landscape at $\varepsilon_{\text{app}} = 8.4\%$ as a function of bond length.
            (b) Energy landscapes at selected $\varepsilon_{\text{app}}$ (vertical gray lines in (a)). The dashed line $-f_{\text{b}}\ell$ is the common tangent between the two minima of the brown curve at $\varepsilon_{\text{app}} = 10.1\%$.
            (c) Energy landscapes after subtraction of the tangent contribution, such that the two minima lie at the same energy.
            (d) Universal barrier shape function $E_{\mathrm{univ}}(\ell^*)$ for a-C across all investigated densities.
            (e) $e_0$ versus $\Delta \ell_0$ for bonds sampled in the plastic flow regime of a-C at $2.7~\text{g}\,\text{cm}^{-3}$. The solid line indicates a cubic scaling with prefactor $0.192~\text{eV}~\text{\AA}^{-3}$.
            (f) $\alpha_0$ versus number density $n$, showing two regimes: ductile (low density) and brittle (high density). In each regime, $\alpha_0$ varies linearly with $n$ (solid blue lines). Insets illustrate the two failure modes; colors represent $D^2_{\min}$~(nonaffine displacement; see Supplementary Note~S-II) at $\varepsilon_{\text{app}}=1$ relative to $\varepsilon_{\text{app}}=0.5$.
            (g) $\Delta \ell$ as a function of $\Delta \ell_0$ for the same bonds as in (e); solid lines are linear fits.}
    \label{fig:single-bond}
\end{figure*}

The well-defined network structure of a-C allows us to disentangle individual STs, revealing them to be single bond-breaking or bond-forming events (Supplementary Notes~S-II and~S-III, also Ref.~\cite{Birks2026}).
The bond-breaking event shown in Fig.~\ref{fig:single-bond}a is an example of an instability with a single ST.
As strain increases, the bond stretches smoothly until $\varepsilon_{\text{app}} \approx 10.1\%$, where it abruptly jumps from about $2.0$ to $2.5$ Å.
Upon reversing the strain (red curve), the bond remains open until $\varepsilon_{\text{app}} \approx 7.3\%$, where it snaps back to the closed state.
This hysteresis reflects the presence of an energy barrier separating the open and closed bond configurations.
A natural reaction coordinate for this process is the bond length $\ell$.
By constraining $\ell$ to specific values and minimizing the system’s potential energy, we obtain the energy profile $E_{\text{pot}}(\ell;\varepsilon_{\text{app}})$ (inset, Fig.~\ref{fig:single-bond}a).
Transition path sampling with the nudged elastic band method~\cite{henkelman_improved_2000,Henkelman2000CI} confirms these results (Supplementary Note~S-IV).

Figure~\ref{fig:single-bond}b shows the energy profiles of the bond discussed above at several applied strains.
The landscapes feature two minima separated by a barrier $E_{\text{b}}(\varepsilon_{\text{app}})$.
At the limits of the hysteresis loop, this barrier vanishes, destabilizing either the closed state ($\varepsilon_{\text{app}} \approx 10.1\%$) or the open state ($\varepsilon_{\text{app}} \approx 7.3\%$).
The bond then jumps by the bond-jump distance $\Delta \ell$, dissipating an
energy $e_{\text{diss}}$ as it relaxes into the new minimum.
Both quantities are marked for bond breaking in Fig.~\ref{fig:single-bond}b
(brown line); bond formation is analogous.

We model each bond as having an intrinsic energy landscape $E_{\text{bond}}(\ell)$ with two equal-energy minima separated by a barrier $e_0$.
Under shear, an effective force $f_{\text{b}}$ acts on the bond, shifting this barrier up or down.
This picture follows the classic Eyring–Kauzmann reaction-rate model~\cite{Eyring1936,kauzmann_viscous_1940}.
During quasistatic shear, the system always resides in a local minimum, so the actual force on an individual bond cannot be measured directly.
Instead, we determine the force using a common tangent to the two minima (dashed brown line in Fig.~\ref{fig:single-bond}b), which represents the work $-f_{\text{b}}\ell$ applied to shift the bond from its intrinsic state.
The effective energy landscape is then given by
\begin{equation}
E_{\text{pot}}(\ell;\varepsilon_{\text{app}})=E_{\text{bond}}(\ell)-f_{\text{b}}(\varepsilon_{\text{app}})\ell.
\end{equation}
Subtracting the respective tangent from the measured landscapes (Fig.~\ref{fig:single-bond}b) collapses them onto a single curve $E_{\text{bond}}(\ell)$ (Fig.~\ref{fig:single-bond}c).
From this, we extract the bond's intrinsic energy barrier $e_0$ and intrinsic bond-jump distance $\Delta \ell_0$, both indicated in Fig.~\ref{fig:single-bond}c.

We conduct a similar analysis on an ensemble of plastic events in the plastic flow regime, each consisting of a single ST.
For each bond, we obtain its energy landscape $E_{\mathrm{bond}}(\ell)$, where $e_0$ and $\Delta \ell_0$ provide natural energy and length scales.
Although each bond resides in a different local environment, rescaling each landscape by these quantities collapses all curves onto a single universal curve $E_{\mathrm{univ}}(\ell^*)$, where $\ell^*=\ell/\Delta\ell_0$ is the dimensionless bond length (Fig.~\ref{fig:single-bond}d).
Remarkably, the collapse extends to every a-C network across the entire density range investigated.
This universality reflects the common microscopic origin of these events: the rehybridization of an $\mathrm{sp}^3$ site toward $\mathrm{sp}^2$ during bond breaking and the reverse process during bond formation~\cite{hybridization_1,hybridization_2}.
The surrounding chemical environment therefore sets $e_0$ and $\Delta \ell_0$, while leaving the dimensionless shape unchanged.

Having established a universal shape, we next examine the relationship between the characteristic scales.
For both bond-breaking and bond-forming events, we find that the intrinsic barrier scales cubically with $\Delta\ell_0$, namely $e_0 = \alpha_0 \, \Delta \ell_0^{\,3}$.
Figure~\ref{fig:single-bond}e shows this for a representative density of
$2.7~\mathrm{g\,cm^{-3}}$, where $\alpha_0=0.192~\text{eV}\,\text{\AA}^{-3}$.
The cubic exponent persists at every density investigated.
Only $\alpha_0$ changes, increasing monotonically with density in two distinct regimes (Fig.~\ref{fig:single-bond}f).
Linear fits of $\alpha_0$ versus number density $n$ within each regime give a weak slope of $0.57~\text{eV}$ at low densities and
a considerably steeper slope of $2.62~\text{eV}$ at high densities.
We find that these regimes correspond to distinct modes of mechanical failure: ductile
deformation at low densities, and brittle yielding accompanied by shear-band
formation at high densities (Supplementary Note~S-V).

The bond-specific energy landscape $E_\mathrm{bond}(\ell)$ can therefore be written for a given density as
\begin{equation}
    E_\mathrm{bond} \left( \ell \right)=\alpha_0 \, \Delta \ell_0^{\,3} \; E_\mathrm{univ} \left(\frac{\ell}{\Delta \ell_0} \right) \; .
    \label{eq:E_bond}
\end{equation}
$E_{\mathrm{univ}}(\ell^*)$ is markedly asymmetric, with the two minima differing in
position and in local curvature about the barrier.
We therefore fit its two branches separately with quartic potentials (Supplementary Note~S-VI), shown as the black curve in Fig.~\ref{fig:single-bond}d.
This analytical representation allows us to locate the saddle points of the landscape and quantify the associated energy dissipation.
The resulting ratio $\chi=e_{\mathrm{diss}}/e_0$ takes the values $\chi_{\mathrm{b}}=3.37$ for bond breaking and $\chi_{\mathrm{f}}=2.64$ for bond formation (Supplementary Note~S-VII).
This implies that $e_{\mathrm{diss}}$ follows the same cubic power-law scaling as the intrinsic barrier, $e_{\mathrm{diss}}=\beta\,\Delta\ell_0^3$, with prefactors $\beta_{\mathrm{b}}=\chi_{\mathrm{b}}\alpha_0$ for bond breaking and $\beta_{\mathrm{f}}=\chi_{\mathrm{f}}\alpha_0$ for bond forming.

\begin{figure*}
    \centering
    \includegraphics[]{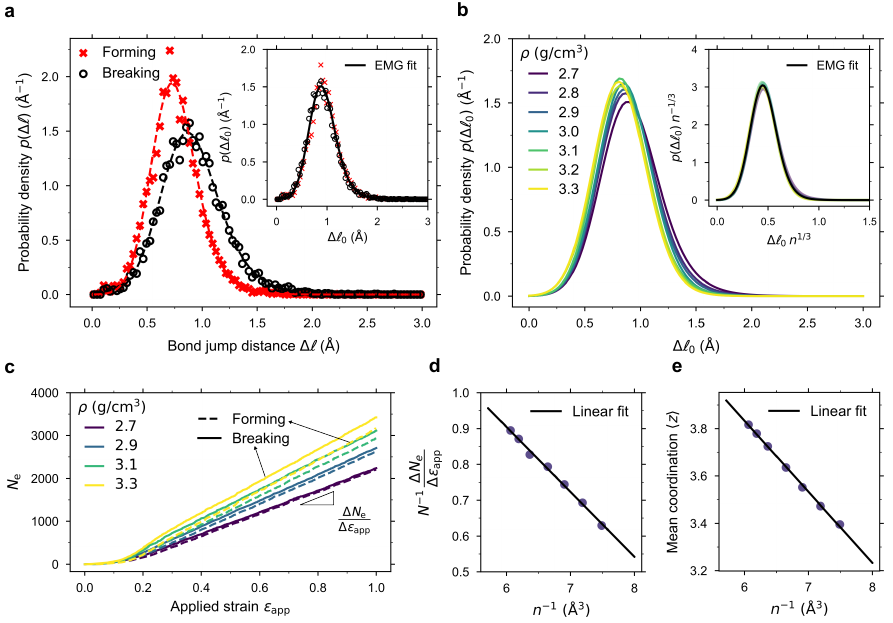}
    \caption{{\bf Statistics of shear transformations in a-C.}
    (a) $p(\Delta\ell)$ for bond-breaking (black circles; $p_\mathrm{b}(\Delta\ell)$) and
    bond-forming (red crosses; $p_\mathrm{f}(\Delta\ell)$) events in the plastic flow regime at $2.7~\text{g}\,\text{cm}^{-3}$. Dashed lines are guides to the eye.
    The coordinate transformation $\Delta \ell \rightarrow \Delta \ell_0$ maps both distributions
    onto a single curve $p(\Delta \ell_0)$, shown in the inset with an exponentially modified Gaussian (EMG) fit as a solid black line.
    (b) $p(\Delta \ell_0)$ for all investigated densities. The further transformation
    $\Delta \ell_0^* = \Delta \ell_0\, n^{1/3}$ yields a single
    density-independent master curve (inset), also well described by an EMG fit
    $\left(\mu_\mathrm{p}=0.388,\sigma_\mathrm{p}=0.114,\lambda_\mathrm{p}=13.46\right)$.
    (c) Cumulative number of broken (solid lines) and formed (dashed lines) bonds versus
    $\varepsilon_\mathrm{app}$ for different densities. In the plastic flow regime, both occur at the same rate for a given density, as reflected
    by the similar slopes $\Delta N_\mathrm{e} / \Delta \varepsilon_\mathrm{app}$.
    (d) $\Delta N_\mathrm{e} / \Delta \varepsilon_\mathrm{app}$ per atom in the steady-state flow regime as a function of the inverse number density $n^{-1}$, showing a clear linear dependence.
    (e) Mean coordination number $\langle z \rangle$ as a function of $n^{-1}$, exhibiting also a linear dependence.
    }
    \label{fig:bond-statistics}
\end{figure*}

The preceding analysis reveals that the energetics of an ST are fully determined by the intrinsic length scale $\Delta\ell_0$.
However, $\Delta\ell_0$ is an idealized quantity requiring the full energy landscape of each bond, which is impractical for the thousands of STs encountered in the flow regime.
The distance a bond actually jumps when it becomes unstable, $\Delta\ell$, is instead directly observable (Fig.~\ref{fig:single-bond}a).
We find that $\Delta\ell$ depends linearly on $\Delta\ell_0$ (Fig.~\ref{fig:single-bond}g), with slopes of $1.0$ for bond-breaking events and $0.8$ for bond-forming events.
This linear dependence is a direct consequence of the universal barrier shape, while the difference in slope reflects the asymmetry between its two minima.
We can therefore infer $\Delta\ell_0$ for every ST from its observed jump distance.
This naturally raises the question of how $\Delta\ell$ and consequently $\Delta\ell_0$ are distributed.

To determine these distributions, we collect all STs occurring throughout the plastic flow regime, $\varepsilon_{\mathrm{app}}\in[0.5,1]$.
Individual plastic events need not consist of a single ST, as several STs can be triggered collectively within one strain increment as part of an avalanche~\cite{sethna_crackling_2001}.
We therefore identify all bonds that break or form within each quasistatic step using the screening criterion of Supplementary Note~S-III and count each as a separate ST.
We exclude from the statistics STs in which a single atom participates in several events within the
same avalanche, since the jump distance is ill-defined.
Figure~\ref{fig:bond-statistics}a shows the distributions $p_\mathrm{b}(\Delta\ell)$ and $p_\mathrm{f}(\Delta\ell)$ for bond-breaking and bond-forming events, respectively.
The two distributions are distinct, with $p_\mathrm{b}(\Delta\ell)$ biased toward larger jump distances.

Transforming to the intrinsic coordinate $\Delta\ell_0$ collapses the two distributions onto a single curve $p(\Delta\ell_0)$ (inset, Fig.~\ref{fig:bond-statistics}a), which is well described by an exponentially modified Gaussian (EMG),
\begin{equation}
\begin{aligned}
&p(\Delta\ell_0;\mu_\mathrm{p},\sigma_\mathrm{p},\lambda_\mathrm{p})
\\
&\quad = \frac{\lambda_\mathrm{p}}{2}
\exp\!\left[
    \lambda_\mathrm{p}(\mu_\mathrm{p}-\Delta\ell_0)
    + \frac{\lambda_\mathrm{p}^{2}\sigma_\mathrm{p}^{2}}{2}
\right]
\\
&\qquad {}\times \operatorname{erfc}\!\left[
    \frac{\mu_\mathrm{p}+\lambda_\mathrm{p}\sigma_\mathrm{p}^{2}
          -\Delta\ell_0}
         {\sqrt{2}\,\sigma_\mathrm{p}}
\right].
\end{aligned}
\label{eq:emg}
\end{equation}
where $\mu_\mathrm{p}$, $\sigma_\mathrm{p}$, and $\lambda_\mathrm{p}$ are the location, width, and rate parameters of the EMG.
Figure~\ref{fig:bond-statistics}b presents $p(\Delta\ell_0)$ for systems at different densities.
With increasing density, the distributions shift systematically toward smaller values of $\Delta\ell_0$, suggesting that the relevant length scale is set by the interatomic spacing.
Nondimensionalizing with the number density $n$, whose inverse cube root
$n^{-1/3}$ sets that scale, $\Delta\ell_0^* = \Delta\ell_0\, n^{1/3}$, collapses
all distributions onto a single universal curve $p^*(\Delta\ell_0^*)$ (inset,
Fig.~\ref{fig:bond-statistics}b), again well described by an EMG.

Each bond-breaking event dissipates an energy
$e_{\mathrm{diss}} = \beta_\mathrm{b}\,\Delta\ell_0^{\,3}$, so the average over the
flow regime follows by weighting with $p(\Delta\ell_0)$,
\begin{equation}
\begin{aligned}
\langle e_{\mathrm{diss}} \rangle_\mathrm{b}
&= \beta_\mathrm{b} \int_0^{\infty}
\mathrm{d}\Delta \ell_0 \,
p(\Delta \ell_0)\, \Delta \ell_0^{\,3}
\\
&= \beta_\mathrm{b} \, n^{-1} \int_0^{\infty}
\mathrm{d}\Delta \ell_0^* \,
p^*(\Delta \ell_0^*)\, \Delta \ell_0^{*\,3}.
\end{aligned}
\label{eq:ediss_breaking}
\end{equation}
The second equality follows from transforming the integration variable from
$\Delta\ell_0$ to $\Delta\ell_0^*$.
Bond formation is analogous, with prefactor $\beta_\mathrm{f}$.

We next examine the rate at which STs are activated per unit strain.
Figure~\ref{fig:bond-statistics}c shows the cumulative number of bond-breaking and
bond-forming events as a function of $\varepsilon_\mathrm{app}$
for different densities.
Bond breaking initially occurs at a slightly higher rate than bond formation, but
beyond $\varepsilon_\mathrm{app}=0.5$ (steady-state flow) the two rates converge at
every density.
This balance reflects the structural invariance of the network under continued
deformation, which supports a steady flow stress.
Dividing the steady-state slopes by the number of atoms $N$ gives the ST activation
rate per atom, $N^{-1}\Delta N_\mathrm{e}/\Delta \varepsilon_\mathrm{app}$.
This rate is linear in the inverse number density
(Fig.~\ref{fig:bond-statistics}d), with an intercept of $2$ STs per atom and unit
strain and a slope of $-0.18~\text{\AA}^{-3}$.
We have no quantitative model for this scaling, but the mean coordination number
$\langle z \rangle$ is also linear in $n^{-1}$ (Fig.~\ref{fig:bond-statistics}e).
This suggests that network connectivity controls the activation rate.
More highly coordinated networks face stricter compatibility constraints, and must
break and form more bonds to accommodate the same plastic strain.

\begin{figure}
\centering
    \includegraphics[]{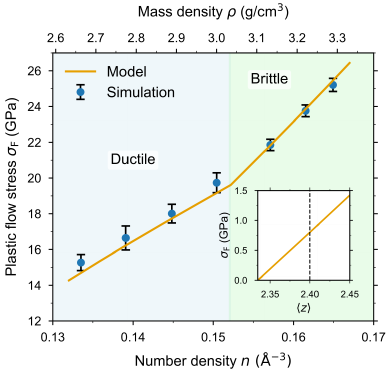}
    \caption{\textbf{Prediction of the plastic flow stress.}
    Plastic flow stress $\sigma_{\mathrm{F}}$ versus density from the microscopic prediction, Eq.~\eqref{eq:sigma_f_microscopic} (solid orange line) and macroscopic energy dissipation rates (filled blue circles, with error bars showing standard deviations across ten independent realizations per density).
    Shading marks the ductile (low-density) and brittle (high-density) deformation regimes identified in Fig.~\ref{fig:single-bond}f.
    The inset shows $\sigma_{\mathrm{F}}$ vanishing at mean coordination $\langle z \rangle \approx 2.33$; the vertical dashed line marks the network-glass rigidity-percolation threshold at $\langle z \rangle \approx 2.4$.}   
    \label{fig:prediction}
\end{figure}

We now have all the components needed to calculate the energy dissipated per unit strain.
Normalized by the system volume $V$, this quantity is the plastic contribution to the stress and, in steady state, equals the plastic flow stress
$\sigma_\mathrm{F}$.
This neglects the $\sim 4\%$ of the supplied work stored rather than dissipated (Fig.~\ref{fig:phenomenology}b), which lies within the statistical uncertainty of our data (Fig.~\ref{fig:prediction}).
At a given density, the flow stress is therefore
\begin{equation}
    \sigma_{\text{F}} = \frac{1}{V} \frac{\mathrm{d} E_{\text{diss}}}{\mathrm{d} \varepsilon_\mathrm{app}} = 
    \frac{1}{V} \frac{\Delta N_{\text{e}}}{\Delta \varepsilon_\mathrm{app}} \left(  \langle e_\mathrm{diss} \rangle_\mathrm{b} + \langle e_\mathrm{diss} \rangle_\mathrm{f} \right).
    \label{eq:sigma_f_macroscopic}
\end{equation}
Inserting Eq.~\eqref{eq:ediss_breaking} and its bond-forming counterpart gives a
microscopic model for the flow stress across all densities,
\begin{equation}
\begin{aligned}
\sigma_{\mathrm{F}}(n)
&=
\frac{N^{-1}\Delta N_{\mathrm{e}}}
     {\Delta \varepsilon_\mathrm{app}}(n)\,
\alpha_0(n)
\\
&\quad {}\times
\left[
(\chi_{\text{b}} + \chi_{\text{f}})
\int_0^{\infty}
\mathrm{d}\Delta \ell_0^*\,
p^*(\Delta \ell_0^*)\,
\Delta \ell_0^{*\,3}
\right].
\end{aligned}
\label{eq:sigma_f_microscopic}
\end{equation}
The bracketed term is a numerical constant with a value $\approx 0.75$.
In Supplementary Note~S-VIII, we present a detailed analysis supporting the system-size independence of all terms in Eq.~\eqref{eq:sigma_f_microscopic}.
Figure~\ref{fig:prediction} compares the resulting prediction of Eq.~\eqref{eq:sigma_f_microscopic} with the plastic flow stress obtained independently from the macroscopic energy dissipation.
The agreement is quantitative across the full density range, including the high-density regime where deformation localizes into shear bands (Fig.~\ref{fig:single-bond}f).
Interestingly, the model extrapolates to a vanishing flow stress at $\langle z \rangle \approx 2.33$ (inset of Fig.~\ref{fig:prediction}), close to the rigidity-percolation threshold of network glasses at $\langle z \rangle \approx 2.4$~\cite{Thorpe_1985}.

Our central result is that the plastic flow stress of a-C follows from
independently determined microscopic quantities.
This rests on the striking simplicity of STs in a-C, single bond-breaking or
bond-forming events described by one reaction coordinate, the bond length.
Their entire energetics follows from one observable, the bond-jump distance.
Our framework thus provides a microscopic foundation for phenomenological ST
models~\cite{Albaret_2016,Zhang_2022}.
The construction should extend to any network glass with individually resolvable
bonds.
Additionally, because the energy landscape sets the scale for thermally activated bond
switching, it opens a route to temperature- and strain-rate-dependent plasticity.

\section*{Acknowledgments}

We thank Fraser Birks, Maciej Buze and James Kermode for useful discussions.
We are indebted to Gabor Cs\'anyi and Volker Deringer for providing an early version of the a-C Gaussian approximation potential which we used in the initial phases of this project, and Matous Mrovec for providing an early version of the a-C ACE potential which we used for the final calculations. This research was partially supported by the Deutsche Forschungsgemeinschaft (Research Unit 5099, grant PA 2023/6). Computations were carried out on NEMO (University of Freiburg, DFG grant INST 39/963-1).

\section*{Author contributions}

I.G. and R.J. performed the simulations and carried out the analysis. W.G.N., M.M. and L.P. conceptualized the research. I.G., R.J. and L.P. wrote the manuscript. All authors discussed the results and commented on the manuscript.

\section*{Software}
Simulations were performed with \texttt{LAMMPS}~\cite{Thompson2022} and \texttt{ASE}~\cite{HjorthLarsen2017}. \texttt{OVITO}~\cite{Stukowski2010} was used for analysis and visualization.

% Bibliography of the main text (inlined from BibTeX/apsrev4-1 output)

\clearpage

%%%%%%%%%%%%%%%% START OF SUPPLEMENT %%%%%%%%%%%%%%%

% Restart SI numbering
\setcounter{section}{0}
\setcounter{figure}{0}
\setcounter{table}{0}
\setcounter{equation}{0}
\setcounter{page}{1}

% SI numbering format
\renewcommand{\thesection}{S-\Roman{section}}
\renewcommand{\thefigure}{S\arabic{figure}}
\renewcommand{\thetable}{S\arabic{table}}
\renewcommand{\theequation}{S\arabic{equation}}
\renewcommand{\thepage}{S\arabic{page}}

\setcounter{secnumdepth}{4}

%%%%%%%%%%%%%%%% SUPPLEMENT TITLE PAGE %%%%%%%%%%%%%%%
\onecolumngrid
\begin{center}

\section*{Supplementary Information for ``Universal barrier shape for plastic flow in a network glass''}

Ibrahim~Ghanem,
Richard~Jana,
Wolfram~G.~N\"ohring,
Michael Moseler,
Lars~Pastewka$^{\ast}$\\
\small$^\ast$Corresponding author. Email: lars.pastewka@imtek.uni-freiburg.de
\end{center}

\twocolumngrid

\section{Liquid-quench and shearing protocols}

We used molecular dynamics simulations for the preparation of a-C networks.
Carbon atoms interacted via a machine-learned atomic cluster expansion potential (ACE), similar to the work of Ref.~\cite{Qamar2023-nw} but without fitting dispersion corrections into the potential.
This specific potential yields energy landscapes that are smooth enough for the barrier calculations shown in this paper not to become trapped in spurious minima.
Amorphous structures ($4096$ atoms) were generated through a liquid-quench process, starting from a diamond configuration  heated to $10{,}000~\text{K}$.
The resulting melt was equilibrated at this temperature for $100~\text{ps}$ before being rapidly quenched to $0~\text{K}$ at a rate of $1000~\text{K}\,\text{ps}^{-1}$.
Following the quench, the system underwent further optimization, including relaxation of the triclinic simulation box, to achieve stress-free structures.
The optimizations ran until the norm of the global force vector dropped below $10^{-8}~\text{eV}\,\text{\AA}^{-1}$.
Using this methodology, we generated a-C networks of varying mass densities $\rho$ above $2.7~\text{g}\,\text{cm}^{-3}$, producing ten independent realizations for each density. 
All statistical data were obtained from these realizations.

\begin{figure*}[ht]
\centering
    \includegraphics[]{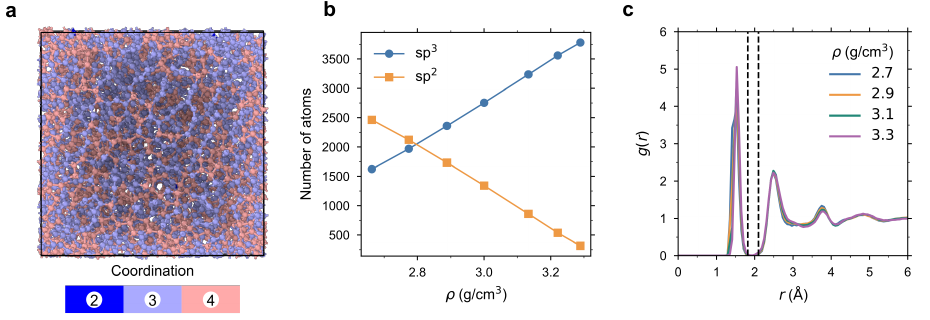}
    \caption{{\bf Initial structures.} (a) Snapshot of a stress-free a-C network with a density of $2.7~\text{g}\,\text{cm}^{-3}$, with atoms colored by coordination number. (b) Number of sp$^3$ and sp$^2$ atoms as a function of density. (c) Radial distribution function $g(r)$ for different densities. The dashed vertical lines mark the region between the first and second neighbor shells where $g(r)$ vanishes.}
    \label{fig:S_initial_structure}
\end{figure*}

Figure~\ref{fig:S_initial_structure}a shows an a-C network with a density of $2.7~\text{g}\,\text{cm}^{-3}$.
Most atoms are three- or fourfold coordinated, with rare occurrences of twofold coordinated atoms.
The sp$^3$ content increases with density at the expense of sp$^2$ content, as shown in Fig.~\ref{fig:S_initial_structure}b.
Figure~\ref{fig:S_initial_structure}c shows the radial distribution function $g(r)$ for different densities.
For all densities, $g(r)$ vanishes between the first and second neighbor shells, as indicated by the two dashed vertical lines.

To probe the plastic response, we subjected the systems to simple shear by deforming the periodic simulation cell in incremental strain steps of $\Delta \varepsilon_{\mathrm{app}} = 0.02\%$.
The deformation was applied in the $xy$-plane, with the strain gradient oriented along the $y$-axis.
After each strain increment, the systems were relaxed to an energy-minimized configuration using a global force tolerance of $10^{-8}~\text{eV}\,\text{\AA}^{-1}$, resulting in an athermal quasistatic deformation protocol.

\section{Bond breaking and forming}

\begin{figure*}[ht]
\centering
    \includegraphics[]{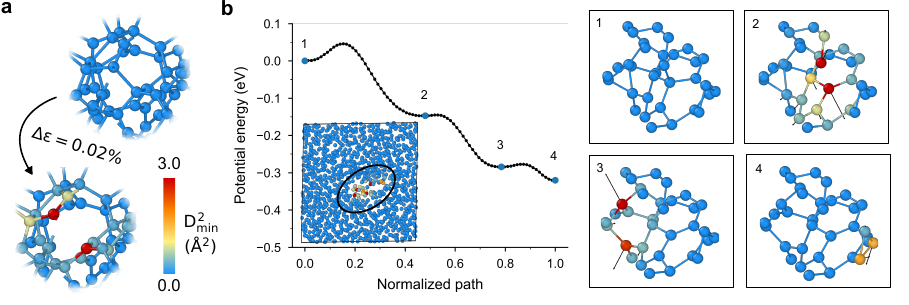}
    \caption{{\bf Shear transformations in a-C.} (a) Bond-breaking event representing an elementary shear transformation in a-C. The color code represents the $D^2_\mathrm{min}$ which peaks right after the plastic event and is localized on the two atoms participating in the event. (b) MEP of an avalanche computed between two stable configurations at the same applied strain, before and after the instability. The avalanche is detected during quasistatic shear by a pronounced peak in the $D^2_\mathrm{min}$ field (inset). On the right, we display the $D^2_\mathrm{min}$ and atomic displacement fields associated with the transitions between successive minima along the MEP (zoomed to the avalanche region), demonstrating that each transition corresponds to the breaking or formation of a covalent bond.}
    \label{fig:S_single_event}
\end{figure*}

We identify plastic events in our systems through a microscopic analysis of atomic rearrangements using the $D^2_\mathrm{min}$ metric of Falk and Langer~\cite{falk_dynamics_1998}. 
A detailed examination shows that these events consistently involve abrupt jumps between the first and second neighbor shells.
In other words, all plastic nonaffine motions in our a-C networks correspond to the breaking or formation of covalent bonds.
Figure~\ref{fig:S_single_event}a shows an example of a quasistatic frame involving the breaking of a single bond.
The $D^2_\mathrm{min}$ value peaks right after the plastic event and is localized on the two atoms participating in the broken bond.
While there is elastic relaxation of the surrounding medium, the core irreversible process is the breaking of the bond.

We extend this analysis to avalanches in a-C and continue to use $D^2_{\mathrm{min}}$ as the reference indicator for plastic activity.
To analyze the constituents of an avalanche, we consider two configurations at the same applied strain, before and after the instability.
Both configurations are obtained by shearing the system back in the opposite direction until the desired strain is reached.
We then perform a Climbing Image Nudged Elastic Band (CI-NEB)~\cite{henkelman_improved_2000,Henkelman2000CI} calculation between the two states.
We use a spring constant of $k = 10~\text{eV}\,\text{\AA}^{-2}$ and set the force convergence criterion to $\text{ftol}=10^{-4}~\text{eV}\,\text{\AA}^{-1}$.
The resulting minimum energy path (MEP), shown in Fig.~\ref{fig:S_single_event}b, consists of multiple minima separated by energy barriers. 
The minima (highlighted by blue dots) correspond to stable configurations along the MEP.
We compute the $D^2_{\mathrm{min}}$ field and atomic displacement field for each minimum, using the preceding minimum along the path as the reference configuration.
As shown in the zoomed-in snapshots in Fig.~\ref{fig:S_single_event}b, each transition between consecutive minima corresponds to the breaking or formation of a covalent bond, and each elementary event closely resembles the single bond-breaking event shown in Fig.~\ref{fig:S_single_event}a.

These observations lead us to propose that, in a-C networks, a shear transformation corresponds to the breaking or formation of a single bond. 
Within this framework, avalanches can be interpreted as plastic events composed of a chain of bond-breaking and bond-forming processes.

\section{Detecting bond breaking and formation events}

Bond jumps occur for a bond between atoms $i$ and $j$ with distance $\ell_{ij}$ within the region where $g(r)$ is nearly zero (see Fig.~\ref{fig:S_initial_structure}c).
A naive approach to determine whether bond $i$-$j$ is broken would be to compare $\ell_{ij}$ with some cutoff distance $\ell_\mathrm{c}$ within this region.
This assumes that \emph{all} formed bonds have $\ell_{ij}<\ell_\mathrm{c}$ and \emph{all} broken bonds have $\ell_{ij}>\ell_\mathrm{c}$.
However, we observe that each bond crosses a different section of the ``forbidden'' region of the $g(r)$, such that no unique $\ell_\mathrm{c}$ exists that captures the state of all bonds in the system. 

\begin{figure*}[ht]
\centering    
\includegraphics[]{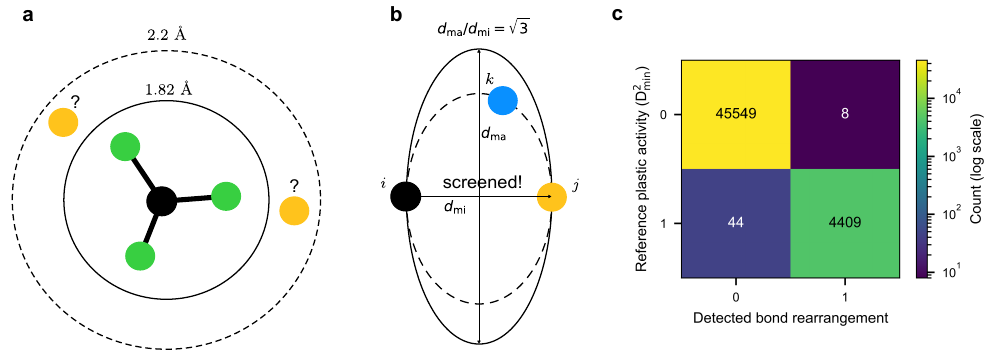}
    \caption{{\bf Screening-augmented neighbor list analysis.} (a) Atoms are considered bonded to a reference atom if they lie within a cutoff distance of $1.82~\text{\AA}$. Atoms located between $1.82~\text{\AA}$ and $2.2~\text{\AA}$ are subject to a screening criterion to determine their bonding status. (b) To evaluate whether a bond between atoms $i$ and $j$ is screened, an ellipsoid is constructed with its minor axis aligned along the bond and an axis ratio of $\sqrt{3}$. If a third atom lies within this ellipsoid, the bond is classified as screened and thus excluded. (c) Confusion matrix quantifying the agreement between reference plastic activity identified using $D^2_{\mathrm{min}}$ and plastic events detected by our bond-rearrangement algorithm.}
    \label{fig:S_screening}
\end{figure*}

We therefore employ a different criterion to determine neighbor relationships, inspired by the idea of bond screening.
We consider a bond $i$-$j$ formed or broken if the distance $\ell_{ij}$ is clearly outside of the region where $g(r)$ is zero, $\ell_{ij}<\ell_\mathrm{min}=1.82~\text{\AA}$ for formed bonds and
$\ell_{ij}>\ell_\mathrm{max}=2.2~\text{\AA}$ for broken bonds (see Fig.~\ref{fig:S_screening}a).
In the intermediate region, we consider it formed if no third atom exists within the neighborhood of the bond.
The specific criterion we use was introduced by Baskes, Angelo and Bisson~\cite{Baskes_Angelo_Bisson_1994} in the context of embedded-atom method potentials, who used a rotational ellipsoid with minor axis centered on the bond to define what the neighborhood of a bond is. 
We consider bonds screened or broken where a third atom sits within the rotational ellipsoid with a ratio of $\sqrt{3}$ between main axis and other axes as depicted in Fig.~\ref{fig:S_screening}b (see also \cite{Pastewka_Klemenz_Gumbsch_Moseler_2013}).

We assess the accuracy of our bond-rearrangement detection algorithm by benchmarking it against the nonaffine displacement measure $D^2_{\mathrm{min}}$.
A plastic event is deemed to occur whenever at least one atom exhibits $D^2_{\mathrm{min}} > 0.1~\text{\AA}^2$, which we treat as the reference input signal.
Independently, our algorithm reports a positive detection whenever a covalent bond is broken or formed between two consecutive quasistatic shear configurations, yielding the measured signal.
The correspondence between the two criteria is quantified using the confusion matrix shown in Fig.~\ref{fig:S_screening}c. 
The agreement, with few false positives and false negatives, demonstrates that our algorithm successfully identifies plastic activity in amorphous carbon.

\section{NEB calculations}
\begin{figure}[ht]
\centering \includegraphics[]{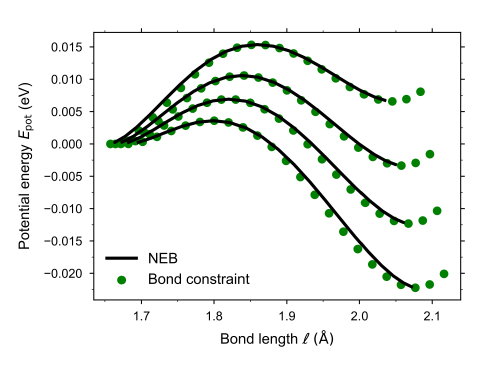}
    \caption{{\bf Bond-constrained optimization vs. NEB.} Energy landscapes of a single bond at various applied strains. Green circles represent the energy calculated using the bond-constrained optimization method. Complementary CI-NEB calculations (black continuous lines) validate this approach by reproducing the same energy barrier profiles.}
    \label{fig:S_NEB}
\end{figure}

In Fig.~2b (main text) we compute the energy landscapes of a single bond at various applied strains.
The simplicity of the reaction coordinate, the bond length, enables us to extract these landscapes by constraining the bond length to fixed values while simultaneously minimizing the energy with respect to all other degrees of freedom in the system.
In these calculations, the convergence criterion requires that no atom experiences a force greater than $10^{-4}~\text{eV}\,\text{\AA}^{-1}$.
An example of such an analysis is shown in Fig.~\ref{fig:S_NEB}, where the distinct scatter plots (green circles) represent the energy landscapes of a bond at various applied strains.

From the previous analysis, we extract atomistic configurations corresponding to the two local minima: the closed configuration (left) and the open configuration (right) at each applied strain.
These configurations are then used as endpoints in a CI-NEB calculation, with a spring constant of $k = 100~\text{eV}\,\text{\AA}^{-2}$ to couple adjacent images.
Each calculation employs 20 intermediate images, and the convergence criterion is set to $\text{ftol}=10^{-4}~\text{eV}\,\text{\AA}^{-1}$, consistent with the threshold used previously.
The results of the NEB analysis are shown in Fig.~\ref{fig:S_NEB} as black continuous lines connecting the two minima of each energy landscape previously obtained via bond-constrained optimization.
The close agreement between the two methods validates the use of bond-constrained optimization for extracting energy landscapes throughout our study.

\section{Quantifying shear localization}
In this study, the systems are subjected to simple shear along the $x$-direction, with the strain gradient oriented along the $y$-direction.
Under these loading conditions, our results show that whenever a system develops a persistent shear band in the steady-state flow regime, the band is invariably oriented perpendicular to the $y$-axis.

To quantify shear localization, we compute the nonaffine deformation field $D^2_{\mathrm{min}}$ at an applied strain $\varepsilon_{\mathrm{app}} = 1.0$, using the configuration at $\varepsilon_{\mathrm{app}} = 0.5$ as a reference.
The resulting $D^2_{\mathrm{min}}$ field is binned along the $y$-direction, and the mean value in each bin is evaluated to obtain a discrete set $\{p_i\}_{i=1}^{M}$, where $p_i$ denotes the bin-averaged plastic activity (mean $D^2_{\mathrm{min}}$).

As a scalar measure of shear localization, we employ the Gini coefficient, defined as
\begin{equation}
G
=
\frac{1}{2 M^2 \bar{p}}
\sum_{i=1}^{M}
\sum_{j=1}^{M}
\left| p_i - p_j \right|,
\qquad
\bar{p} = \frac{1}{M}\sum_{i=1}^{M} p_i ,
\end{equation}
where $M$ is the number of bins.
The Gini coefficient has been adopted in the context of amorphous solids to quantify strain localization~\cite{Gini}.
In the limit of homogeneous deformation, all bins carry identical plastic activity ($p_i=\bar{p}$), such that $G=0$.
Conversely, in the case of perfectly localized deformation corresponding to plastic activity concentrated within a single bin, the Gini coefficient approaches unity, $G\to 1$.

\begin{figure*}[ht]
\centering \includegraphics[]{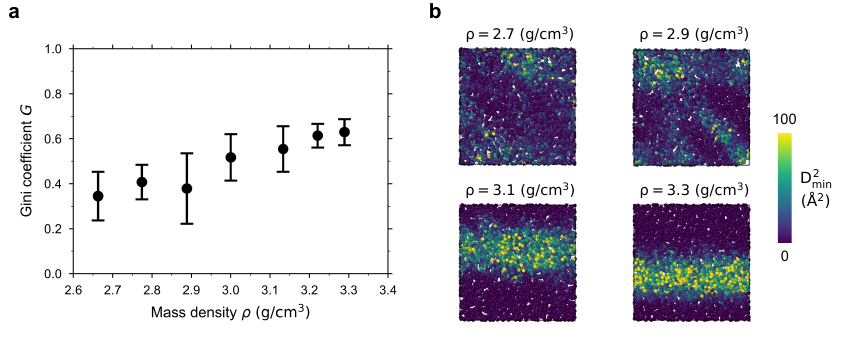}
    \caption{{\bf Shear localization as a function of density.}
    (a) Gini coefficient $G$ as a function of system density. Filled black circles represent averages over ten independent realizations at each density, with error bars indicating the corresponding standard deviation.
    (b) Representative snapshots of selected systems, with atoms color-coded by their $D^2_{\mathrm{min}}$ values. The $D^2_{\mathrm{min}}$ field is computed at $\varepsilon_{\mathrm{app}} = 1.0$ relative to a reference configuration at $\varepsilon_{\mathrm{app}} = 0.5$. For clarity, the systems are mapped onto a rectangular simulation box. The data in both panels are consistent with a progressive localization of plastic activity at higher densities.}
    \label{fig:S_distance_std}
\end{figure*}

Figure~\ref{fig:S_distance_std}a shows the Gini coefficient $G$ as a function of system density.
Filled black circles represent averages over ten independent realizations at each density, with error bars indicating the corresponding standard deviation.
Overall, higher-density a-C networks exhibit larger Gini coefficients, indicating more localized plastic flow in the steady-state regime.
This trend is illustrated in Fig.~\ref{fig:S_distance_std}b, which presents representative snapshots of selected systems, with atoms color-coded according to their $D^2_{\mathrm{min}}$ values.
As the density increases, plastic activity becomes progressively confined to narrow shear bands, consistent with the larger Gini coefficients observed at higher densities in Fig.~\ref{fig:S_distance_std}a.

\section{Fitting $E_\mathrm{univ}(\ell^*)$}
To determine a functional form of $E_\mathrm{univ}(\ell^*)$, we fit each branch (separated by the energy axis) to a quartic polynomial of the form:
\begin{equation}
    f(x; a, b, c) = a x^4 + b x^3 + c x^2 + 1.
    \label{eq:S_quartic}
\end{equation}
The fitting is performed using a least squares approach with the objective function
\begin{equation}
\begin{aligned}
    S(\boldsymbol{\theta}) = {} & \sum_{i=1}^{N} \left[ y_i - f(x_i; \boldsymbol{\theta}) \right]^2 \\
    & + w^2 \left[ f(x_{\min}; \boldsymbol{\theta}) \right]^2
    + w^2 \left[ f'(x_{\min}; \boldsymbol{\theta}) \right]^2,
\end{aligned}
    \label{eq:S_objective_func}
\end{equation}
where \(\boldsymbol{\theta} = (a, b, c)\) is the vector of fitting parameters.
The last two terms in Eq.~\eqref{eq:S_objective_func} act as constraint residuals enforcing that the fitted function has a minimum of zero energy and a vanishing slope at $x_{\min}$.
The position $x_{\min}$ for each branch is obtained by averaging the corresponding minima of the collapsed energy landscapes.
We choose \(w = 1000\) to enforce these constraints.
The fitting process yields separate sets of parameters which are summarized in Table~\ref{tab:S_fit_double_well}.

\setlength{\abovecaptionskip}{10pt}
\begin{table}[htbp]
    \centering
    \caption{Fitting parameters for the left and right branches of the universal
    barrier shape function $E_\mathrm{univ}(\ell^*)$.}
    \label{tab:S_fit_double_well}
    \vspace{6pt} % adds space between caption and first \hline
    \begin{tabular}{l@{\hskip 6pt}@{\hskip 6pt}c@{\hskip 12pt}c@{\hskip 12pt}c}
        \hline
        & $a$ & $b$ & $c$ \\
        \hline
        Left  & 16.1 & -5.3 & -10.6 \\ 
        Right & -4.9 & 18.1 & -11.8 \\ 
        \hline
    \end{tabular}
\end{table}

\section{Scalings of $E_\mathrm{univ}(\ell^*)$}
\begin{figure*}[ht]
\centering \includegraphics[]{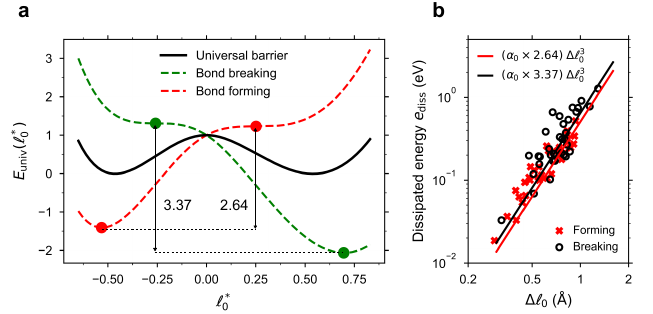}
  \caption{{\bf Scalings of $E_\mathrm{univ}(\ell^*)$.} 
    (a) Piecewise quartic representation of the universal barrier shape
    $E_{\mathrm{univ}}(\ell^*)$ (solid black line). Tilting the reduced landscape by
    a linear term $d x$ leads to a loss of metastability at critical values of $d$,
    corresponding to bond breaking ($d=-3.3$, dashed green line) and bond formation
    ($d=2.8$, dashed red line). The ratios $\chi = e_{\mathrm{diss}}/e_0$ are indicated for both
    processes ($\chi_{\mathrm{b}} = 3.37$ for bond breaking and
    $\chi_{\mathrm{f}} = 2.64$ for bond formation).
    (b) Dissipated energy per ST, $e_{\mathrm{diss}}$, as a
    function of $\Delta \ell_0$ for the same bonds analyzed in Fig.~2e of the main
    text. Solid lines indicate the cubic power-law predictions with prefactors
    $\beta_{\mathrm{b}} = \chi_{\mathrm{b}}\,\alpha_0$ (breaking) and
    $\beta_{\mathrm{f}} = \chi_{\mathrm{f}}\,\alpha_0$ (forming),
    with $\alpha_0 = 0.192~\text{eV}\,\text{\AA}^{-3}$.
    }
    \label{fig:S_Euniv_ediss}
\end{figure*}

We start from the piecewise quartic representation of $E_\mathrm{univ}(\ell^*)$
obtained in Supplementary Note~S-VI, shown as the solid black line in
Fig.~\ref{fig:S_Euniv_ediss}a.
To model the mechanical work performed on an individual bond during deformation,
we add a linear term $d x$, which tilts the reduced landscape.
The resulting tilted function reads
\begin{equation}
f_t(x; a, b, c, d) = f(x; a, b, c) + d x,
\label{eq:S_quartic_tilted}
\end{equation}
where $f(x; a, b, c)$ is given by Eq.~\eqref{eq:S_quartic} with the parameters of
Table~\ref{tab:S_fit_double_well}.
Depending on the sign of $d$, one of the two minima approaches the barrier maximum
as $|d|$ increases, and eventually merges with it in a saddle-node bifurcation.
This loss of metastability signals the onset of a bond-breaking or bond-forming
event.
Solving $f_t' = f_t'' = 0$ numerically, we find that the left (closed) minimum
becomes unstable at $d = -3.3$, corresponding to bond breaking (dashed green line
in Fig.~\ref{fig:S_Euniv_ediss}a), and the right (open) minimum at $d = 2.8$,
corresponding to bond formation (dashed red line).

From the tilted landscapes at the two instability points, we compute the ratio
$\chi = e_{\mathrm{diss}}/e_0$ between the dissipated energy per ST and its intrinsic energy barrier.
As indicated in Fig.~\ref{fig:S_Euniv_ediss}a, we find $\chi_{\mathrm{b}} = 3.37$
for bond breaking and $\chi_{\mathrm{f}} = 2.64$ for bond formation.
This implies that $e_{\mathrm{diss}}$ obeys the same cubic scaling with
$\Delta \ell_0$ as the intrinsic barrier, $e_0 = \alpha_0 \Delta \ell_0^3$
(Fig.~2e of the main text), with prefactors
$\beta_{\mathrm{b}} = \chi_{\mathrm{b}}\alpha_0$ for bond breaking and
$\beta_{\mathrm{f}} = \chi_{\mathrm{f}}\alpha_0$ for bond formation.

We validate this prediction in Fig.~\ref{fig:S_Euniv_ediss}b, where we plot
$e_{\mathrm{diss}}$ against $\Delta \ell_0$ for the same bonds analyzed in
Fig.~2e of the main text.
The system considered has a density of $2.7~\mathrm{g\,cm^{-3}}$ and an intrinsic
prefactor $\alpha_0 = 0.192~\text{eV}~\text{\AA}^{-3}$.
The cubic power-law predictions with prefactors $\beta_{\mathrm{b}}$ and
$\beta_{\mathrm{f}}$ are consistent with the numerical data, as indicated by the
solid lines in Fig.~\ref{fig:S_Euniv_ediss}b.

\section{System-size independence}
In the main manuscript, all analyzed systems span different densities but contain the same number of atoms, $N = 4096$.
Here, we assess the system-size independence of the various terms appearing in Eq.~(6) of the main text.

\begin{figure*}[ht]
\centering \includegraphics[]{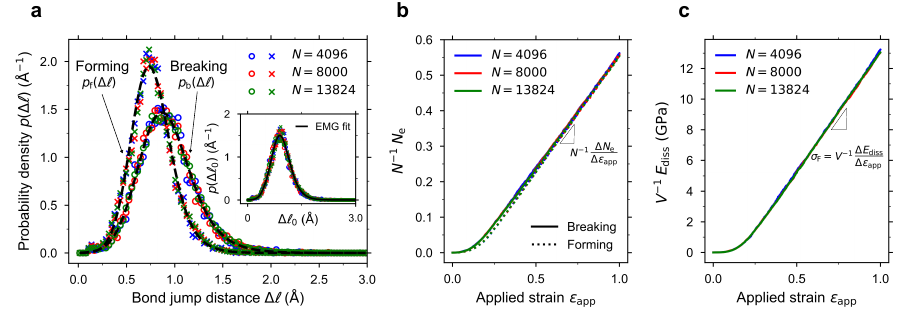}
    \caption{{\bf System-size independence.}
    (a) Probability distributions of bond-jump distances $\Delta \ell$ measured in the steady-state flow regime for a-C systems of different sizes ($N=4096$, $8000$, and $13824$) at fixed density $\rho = 2.7~\mathrm{g\,cm^{-3}}$. No discernible finite-size effects are observed for either bond-breaking or bond-forming events. Upon transformation to the intrinsic coordinate $\Delta \ell_0$, all distributions collapse onto a single master curve (inset).
    (b) Cumulative number of bond-breaking and bond-forming events per atom, $N^{-1} N_{\mathrm{e}}$, as a function of applied strain for different system sizes. In the plastic flow regime, all curves exhibit the same slope.
    (c) Cumulative energy dissipation normalized by volume as a function of applied strain for different system sizes. All curves overlap indicating that beyond $\varepsilon_\mathrm{app} = 0.5$, the plastic flow stress obtained from the slope is system-size independent.
    }
\label{fig:S_N_indep}
\end{figure*}

To this end, we generate additional a-C systems using the same liquid-quench protocol, with larger system sizes $N = 8000$ and $N = 13824$, while keeping the density fixed at $\rho = 2.7~\mathrm{g\,cm^{-3}}$.
Figure~\ref{fig:S_N_indep}a shows the probability distributions of bond-jump distances measured in the steady-state flow regime.
For both bond-breaking and bond-forming events, no discernible finite-size effects are observed.
Moreover, upon performing the coordinate transformation to $\Delta \ell_0$, all distributions corresponding to different system sizes collapse onto a single master curve (inset, Fig.~\ref{fig:S_N_indep}a).
This demonstrates that the rescaled distribution $p^*(\Delta \ell_0^*)$, and consequently the integral appearing in Eq.~(6), is independent of system size.

The rate of shear-transformation activation per atom, $N^{-1}\Delta N_{\mathrm{e}}/\Delta \varepsilon_\mathrm{app}$, is likewise system-size independent, as shown in Fig.~\ref{fig:S_N_indep}b.
Due to the substantial computational cost of the larger systems, we are unable to
generate sufficiently large datasets to directly assess whether the power-law
prefactor $\alpha_0$ or the universal barrier shape $E_\mathrm{univ}(\ell^*)$
depends on system size.
However, Fig.~\ref{fig:S_N_indep}c demonstrates that the plastic flow stress, computed independently from the macroscopic rate of energy dissipation per unit strain, is itself insensitive to system size.

Taken together, these observations imply that either $\alpha_0$ and the ratios $e_{\mathrm{diss}}/e_0$ for bond-breaking and bond-forming events are individually system-size independent, or that their product entering Eq.~(6) is.
Disentangling these two possibilities would require computationally expensive simulations and is therefore left as an open question for future work.

\FloatBarrier
% Bibliography of the SI (inlined; numbering continues from the main text, references cited in both parts are listed only there)

\end{document}